\newcommand{\be}{\begin{equation}}
\newcommand{\ee}{\end{equation}}
\newcommand{\bea}{\begin{eqnarray}}
\newcommand{\eea}{\end{eqnarray}}
\newcommand{\ba}[1]{\begin{array}{#1}}
\newcommand{\ea}{\end{array}}
\documentclass[twocolumn,secnumarabic,amssymb,nobibnotes,nofootinbib,aps,pra,showpacs]{revtex4}

\usepackage{epsfig}
\usepackage{amssymb}
\begin{document}
\setlength{\topmargin}{-0.05in}

\title{Feshbach Resonance Induced   Fano Interference in Photoassociation}
\author{Bimalendu Deb}
\affiliation{Department of Materials Science, and Raman Center for
Atomic, Molecular and Optical Sciences, Indian Association for the
Cultivation of Science (IACS), Jadavpur, Kolkata 700032.}
\author{G. S. Agarwal}
\affiliation{Department of Physics,  Oklahoma State University,
StillWater, OK 74078, USA. }

%\date{\today}

\def\zbf#1{{\bf {#1}}}
\def\bfm#1{\mbox{\boldmath $#1$}}
\def\hf{\frac{1}{2}}
\begin{abstract}

We consider photoassociation from a state of two free atoms when
the continuum state is close to a magnetic field induced Feshbach
resonance and analyze Fano interference in
photoassociation. We show that the minimum in photoassociation
profiles characterized by the Fano asymmetry
parameter $q$ is independent of laser intensity, while the maximum explicitly
depends on laser intensity. We further discuss the possibility of nonlinear Fano effect in photoassociation
near a Feshbach resonance.

\end{abstract}

\maketitle

\section{introduction}

In recent times, quantum interferences have occupied a prominent
place in physics and these occur rather ubiquitously. Many
well-known examples of these include Fano interferences
\cite{Fano61,eberly,gsafano}, electromagnetically induced transparency (EIT)
\cite{EITHarrisPRL89}, vacuum
induced interferences in spontaneous emission \cite{viiGSAbook}. The quantum
interferences have resulted in large number of applications in
coherent control of the optical properties, control of spontaneous
emission \cite{scullyzhuzubairyScience} and slow light
\cite{HauHarris, recentpaperBoyd}. Quantum interference  has been experimentally  demonstrated in coherent formation of molecules  \cite{atom-molecule} and Autler-Townes splitting \cite{two-photon,moal} in two-photon PA.  Theoretical formulation of   PA  within the framework of Fano's theory has  been developed in Refs. \cite{semian} and \cite{semian2}. Recent experimental \cite{Junker:prl:2008, Winkler, ni} and theoretical \cite{Mackie:prl:2008, pellegrini:njp:2009,cote2,Kuznetsova} studies on photoassociation (PA) near a magnetic field Feshbach resonance (MFR) \cite{mfr} have generated a lot of interest in  Fano interference with ultracold atoms. In a remarkable experiment, Junker {\it et al.} \cite{Junker:prl:2008} have demonstrated asymmetric spectral line shape and saturation in PA due to a tunable  MFR. Asymmetric line shape is a hallmark of  Fano-effect and the experimental results of \cite{Junker:prl:2008} can be attributed to the Fano interference.

\begin{figure}
 \includegraphics[width=3.5in]{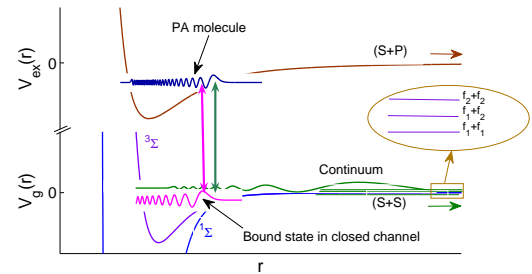}
 \caption{A Schematic diagram showing the diatomic potentials, continuum and bound
 states,  bound-bound and bound-continuum laser couplings. Note that
 the two double-arrow vertical lines refer to the same laser field - green line corresponds to continuum-bound and
the magenta line refers
 to bound-bound couplings. The ground asymptotic channels ($f_a +  f_b$) in the absence of a magnetic field corresponds to
 separated atoms in two different hyperfine numbers $f_a$ and $f_b$.}
 \end{figure}

Here we demonstrate quantum interference in the
context of photoassociation (PA) \cite{pa}  under the
condition when a Feshbach resonance is also involved in
photoassociation. We show Fano like interference minimum in
photoassociation spectrum. In analogy to the well know Fano
q-parameter we can introduce a parameter which governs the
existence of this minimum.  Although the minimum is independent of laser intensity,  the maximum is shown to depend explicitly on laser intensity. From our calculations we extract line shapes which are in broad agreement with the experimental results of junker {\it et al.} \cite{Junker:prl:2008}.
Our formula for photoassociation is expressed in terms of parameters
each of which has a clear physical meaning and is measurable. We derive probability of PA excitation  for arbitrary intensities of the laser field and thus we also discuss nonlinear Fano effect. 
The current work has  some features in common with the recent paper of Kuznetsova {\it et al.} \cite{Kuznetsova} though these authors address a different problem which is the population transfer
using two laser beams. Our emphasis is on quantum interferences in PA using a single
laser beam.

The paper is organized in the following way. In section 2, we consider a simple model of three-channel time-independent scattering in the presence of an optical and a magnetic field.  By using Green's functions, we present compact analytical solution of the model. We then discuss selective results in Sec.3. The paper is concluded in Sec.4.

\section{The model and its solution}

To begin with, we  model PA in the presence  of a Feshbach resonance as a three-channel
scattering problem. There are two ground-state
asymptotic hyperfine channels of which one is closed and the other
one is open. The third channel corresponds to the photoassociated excited molecular configuration. The two ground state channels are  coupled via hyperfine interaction.
At a Feshbach
resonance, the two atoms will form a quasibound state  in the closed channel as schematically illustrated in
Fig.1.  As the strength of the applied magnetic field is
varied, this quasibound state  can
move across the collision energy. When a PA laser is applied to form an
excited photoassociated molecule (PM), there arise  two competing
pathways of dipole transitions as shown by different colors in
Fig.1. One is the continuum-bound and the other one is bound-bound
transition.   We assume that the energy
spacing of closed-channel quasibound  states and  rotational spacing of PM states are
much larger than PA laser line width  so that only one rotational
level ($J$) of a particular vibrational state $v$ of PM is coupled to a
particular quasibound state by the PA laser.

Let us write an energy
eigenstate of the system of two atoms interacting  simultaneously
with a magnetic and a PA laser field in the form
 \bea \mid \Psi_{E} \rangle = \Phi_{f} \mid g_2 \rangle  + \chi \mid g_1 \rangle
  +  \Phi_{p} \mid e \rangle \label{dstate} \eea
where $E$ is an energy eigenvalue, $\mid g_{1(2)} \rangle$
represents the internal electronic states of  1(2) or open(closed)
channel and $\mid e \rangle$ denotes the electronic state of the
excited molecule. $\Phi_{f}$  and $\Phi_{p}$ are the diatomic
bound states. The continuum state  has the form $\chi = \int d
E' b_{E'} \psi_{E'}$ where $\psi_{E'}$  is an energy-normalized
scattering state of collision energy $E'$ and $b_{E'}$ is the
density of unperturbed continuum states. The state (\ref{dstate}) is assumed to be energy-normalized.
The  Hamiltonian of the system can be
written as $ H = H_{kin} + H_{elec} + H_{hfs} + H_B + H_{L}$ where
$H_{kin}$ denotes a term corresponding to the total kinetic energy
of the two atoms and $H_{elec}$ is a term that depends on only
electronic coordinates of the two atoms, $H_{hfs}$ is the
hyperfine interaction term. Here $H_B$ represents the magnetic
interaction in the atomic states, and $H_L$ the laser interaction between
atomic or molecular states. From the time-independent Schr\"{o}dinger
equation $H\Psi_{E} = E \Psi_{E}$ under Born-Oppenheimer
approximation, one obtains the following coupled equations \bea
\left [ -\frac{\hbar^2}{2\mu}\frac{d^2}{d r^2} + B_J(r) \right]
\Phi_{p} &+&
\left ( V_{e}(r) - \hbar \omega_L - E - i \hbar \frac{\gamma}{2} \right ) \Phi_{p}  \nonumber \\
 &=& - \Lambda_1 \chi - \Lambda_2 \Phi_{f} \label{coup1}\eea \be
\left [ -\frac{\hbar^2}{2\mu}\frac{d^2}{d r^2}  + V_{2}(r) - E
\right ] \Phi_{f} = - \Lambda_2^{*} \Phi_{p} - V^{*} \chi
\label{coup2}\ee \bea \left [ - \frac{\hbar^2}{2\mu}\frac{d^2}{d
r^2}  + V_{1}(r) - E \right]\chi = - \Lambda_1^{*} \Phi_{p} - V
\Phi_{f} \label{coup3}\eea where $\omega_L$ is the laser
frequency, $\Lambda_1$ and $\Lambda_2$ are the laser-induced
transition dipole  matrix elements between $\mid e \rangle $ and
$\mid g_1 \rangle $, and between $\mid e \rangle $ and $\mid g_2
\rangle$, respectively. Here $V_{i}$ ( $i \equiv 1, 2$) are the
potentials including hyperfine and Zeeman terms, $V_e$ is the
excited state molecular potential and $V$ stands for 
hyperfine spin coupling between closed channel bound state and continuum states.
Here $B_J(r) = \hbar^2 J(J+1)/(2 \mu r^2)]$ is the rotational term of the excited
state.
Note that for the ground scattering and bound  states we have
considered only the zero rotational state. The zero of energy
scale is taken to be the threshold $E_{th}$ of the open channel 1 and the
energies of the  bound states are measured from this
reference. For two homonuclear atoms, the asymptotic form of the potential
 $V_{e}(r \rightarrow \infty) \sim \hbar \omega_A - C_3/r^3$, where
   $C_3$ is the long-range coefficient of
dipole-dipole interaction between one ground state S-atom and
another excited state P-atom and $\omega_A$ is the atomic frequency.  These three coupled
equations can be solved exactly by the use of real space Green's
function as described below.

\begin{figure}
 \includegraphics[width = 3.5 in]{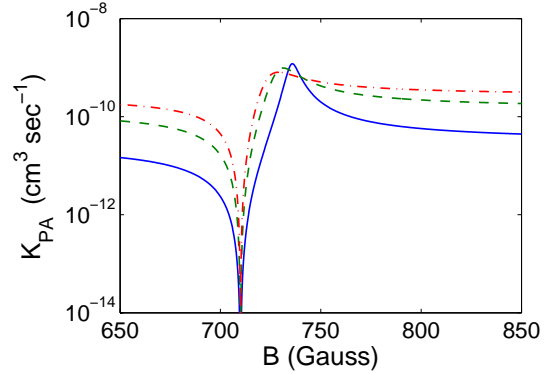}
  \caption{PA rate $K_{PA}$ in cm$^3$ sec$ ^{-1}$ is plotted as a function of magnetic  field $B$  in Gauss. The parameters are $\Gamma_p = 0.1$ MHz (solid line), $\Gamma_p = 0.5 $ MHz (dashed line) and $\Gamma_p = 1.0$ MHz (dashed-dotted line). The parameter $q_{f}$ is chosen to be - 8.90  (solid line), -7.63 (dashed) and  - 6.36 (dashed-dotted) so that the minimum and maximum of $K_{PA}$ occur at $B = 710$ G and $B \simeq 736$ G, respectively. The other parameters are $\gamma$ = 11.7 MHz,
 $S_{p c}$= - 2.8 MHz and temperature $T = 10 \mu$K. The laser
is assumed to be tuned on resonance with the continuum-bound transition.}
 \label{fig:1}
\end{figure}

It is convenient to write  $V_{e} = \hbar \omega_A +
\tilde{V}_{ex}$, where $\tilde{V}_{ex}(r \rightarrow \infty) \sim
-C_3/r^3$.  Let $\phi_{p}$ denote the bound state solution of the
the potential $\tilde{V}_{ex}$ and $E_p$ be the corresponding
bound state (negative) energy. The Green's function for
homogeneous part with $\Lambda_1 = \Lambda_2 = 0$ (i.e. without laser couplings) of Eq.
(\ref{coup1}) can be written as \be G_{p}(r,r')= - \frac{1}{\hbar
\delta + E - E_{p} + i\hbar \gamma/2 }\phi_{p}(r)\phi_{p}(r') \ee
where $\delta = \omega_L - \omega_A$. Using this function, we can
write down the solution of equation (\ref{coup1}) in the form
$\Phi_{p} = A_p \phi_{p} (r) $ where \bea A_p = \frac{ \int dr'
\left [ \Lambda_{1}(r')\chi (r') + \Lambda_2(r') \phi_{p}(r')
\right ] \phi_{p}(r')}{\hbar \delta + E - E_{p} + i\hbar \gamma/2
} \label{ap}\eea Similarly, with the use of Green's function for
the homogeneous part of Eq. (\ref{coup2}), we have $ \Phi_{f}(r) = A_m \phi_{f}(r)$ where \bea
A_m &=& \frac{ \int dr' \left [ V^{*}(r')\chi (r') + A_p
\Lambda_2(r') \phi_{p}(r') \right ] \phi_{f}(r')}{ E - E_{f}
}\label{am} \eea where $\phi_{f}(r)$ is the  wave function and
$E_{f}$ is the energy of bound state in the closed channel in the absence of
laser field. Now, we can express $A_p$ in terms of
integrals involving the continuum state $\chi$ and molecular bound
states $\phi_{p}$ and $\phi_{f}$. Then substituting $\Phi_p$ and
$\Phi_f$ expressed in terms of $A_p$ and $\chi$ in Eq.
(\ref{coup3}) and making
 use of the relation $\chi = \int d E b_{E}
\psi_{E}$, we obtain
 \bea  - \frac{\hbar^2}{2\mu}\frac{d^2 }{d r^2} \psi_{E}(r)
&+& [V_{1}(r) - E ]\psi_{E}(r) = - \Lambda_1^{*}(r)
\tilde{A}_p \phi_{p}(r) \nonumber \\
 &-& \frac{( \tilde{V}_{f c} +
\tilde{A}_p \Lambda_{ p f })}{E- E_{f}} V(r) \phi_{p}(r)
\label{psieq}\eea where \bea \tilde{A}_p = \frac{\tilde{\Lambda}_{
p c} (E - E_{f}) + \Lambda_{ p f} \tilde{V}_{f c}} {{\cal D} (E -
E_{f}) - |\Lambda_{ p f}|^2}.
 \label{tap} \eea
Here ${\cal D } = \hbar \delta + E - E_{p} + i\hbar \gamma/2 $,
$\tilde{V}_{f c} = \int d r \phi_{f}(r) V(r) \psi_{E}(r)$, $
\tilde{\Lambda}_{p c} = \int d r \phi_{p}(r) \Lambda_{1}(r)
\psi_{E}(r)$ and $\tilde{\Lambda}_{2,p f} = \int d r \phi_{p}(r)
\Lambda_{2}(r) \phi_{f}(r)$. Equation (\ref{psieq}) can now be
solved by constructing the Green's function with the scattering
solutions of the homogeneous part (i.e., for $\Lambda_1 = V = 0$).
This Green's function can be written as \bea
{\cal K} (r,r') = - \pi [\psi_{E}^{0 ,reg}(r)\psi_{E}^{0 ,irr}(r') + i \psi_{E}^{0 , reg}(r)\psi_{E}^{0 , reg}(r')],
\nonumber \\
\hspace{2.5cm}  (r'>r)  \nonumber \\
{\cal K} (r,r') = - \pi [\psi_{E}^{0 , reg}(r')\psi_{E}^{0 ,
irr}(r)+i\psi_{E}^{0  ,reg}(r)\psi_{E}^{0 , reg}(r')], \nonumber
\\ \hspace{2.5cm}  (r'<r) \nonumber \eea
where  the regular function $\psi_{E}^{0 , reg}(r)$ vanishes at $r
= 0$ and the irregular solution $\psi_{E}^{0 , irr}(r)$ is defined
by boundary only at $r \rightarrow \infty$. These have the
familiar asymptotic behavior $ \psi_{E}^{0 , reg}(r) \sim
j_{0}\cos\eta_{0} - n_{0}\sin\eta_{0}$ and  $ \psi_{E}^{0 ,
irr}(r) \sim n_{0}\cos\eta_{0} + j_{0}\sin\eta_{0} $, where
$j_{0}$ and $ n_{0}$ are the spherical Bessel and Neumann
functions for $\ell = 0$ and $\eta_{0}$ is the s-wave phase shift
in the absence of laser and magnetic field couplings. Here $E =
\hbar^2 k^2/(2 \mu)$ with $\mu$ being the reduced mass of the two
atoms.  Next, we can express the solution of Eq.
(\ref{psieq}) in the following form \bea \psi_{E} &=& \exp(i
\eta_0) \psi_{E}^{0.reg} + \int d r' {\cal K}(r, r') \left [
\Lambda_1^{*}(r') \tilde{A}_p \phi_{p}(r')
\right. \nonumber \\
 &+& \left. \frac{ \tilde{V}_{f c}^{*} +  \tilde{A}_p
\Lambda_{ p f }}{E - E_{f}} V(r') \phi_{f}(r') \right ].
\label{psi} \eea The stimulated line width of photoassociated
 molecule is given by the Fermi-Golden rule expression $\Gamma_p = 2 \pi |\tilde{\Lambda}_{p
 c}^0|^2/\hbar$   and the Feshbach resonance line width is
$ \Gamma_f = 2 \pi  \mid \tilde{V}_{f c}^{0} \mid^2/\hbar $, where
$\tilde{V}_{f c}^{0} = \int d r \phi_{f}(r) V(r)
\psi_{E}^{0,reg}(r)$ and $ \tilde{\Lambda}_{p c}^{0} = \int d r
\phi_{p}(r) \Lambda_{1}(r) \psi_{E}^{0,reg}(r)$. The Stark energy
shift due to laser coupling of PM state with the continuum is
given by $ S_{p c} = \int \int d r' d r \phi_{p}(r)
\Lambda_{1}^{*}(r) {\rm Re}[{\cal K }(r',r)] \Lambda_{1}
(r')\phi_{p}(r') $. Further, the physics of Feshbach resonance
leads us to introduce the parameter
 \bea V_{p f} = \int \int d r' d r
\phi_{f}(r) V(r){\rm Re}[{\cal K }(r',r)]
\Lambda_{1}^*(r')\phi_{p}(r') \nonumber  \eea which represents
 an effective continuum-mediated magneto-optical coupling between
the two bound states where $
S_{f c} = \int \int d r' d r \phi_{f}(r) V^{*} (r){\rm Re}[{\cal K
}(r',r)] V(r')\phi_{f}(r') $  is the energy shift of the 
closed channel bound state due to its coupling with the continuum. Now
writing $\epsilon = (E - \tilde{E}_{f}) /(\Gamma_f/2) $ with
$\tilde{E}_{f} =  E_{f} + S_{f c}$ being the shifted energy of the
closed-channel bound state, and introducing a parameter \bea q_f =
\frac{\Lambda_{ p f} + V_{p f} }{\pi \tilde{\Lambda}_{ p c}^{0}
\tilde{V}_{f c}^{0}} \label{qf} \eea which we call ``Feshbach
asymmetry parameter", we can express
 \bea \tilde{A}_p =
\frac{ \sqrt{\pi \hbar \Gamma_p/2}}{\hbar \Gamma_f/2}
\left ( \frac{\epsilon + q_f } {\epsilon + i} \right )
\frac{\exp(i\eta_0) }{\Delta_p + i \gamma /\Gamma_f + D_I} \label{ap} \eea
where $\Delta_p = [E - ( E_p - \hbar \delta)]/(\Gamma_f/2) =
\epsilon - [E_p - \hbar \delta - \tilde{E}_f ]/(\Gamma_f/2)$ is
independent of laser intensity and $D_I = (- 2 S_{p c} + i
\Gamma_p )/\Gamma_f
 - (\Gamma_p/\Gamma_f) (q_f - i)^2/(\epsilon + i)$ is a parameter which is proportional
 to laser intensity $I$. In writing the above equation we have assumed that $\tilde{V}_{f
c} ^{0} $, and $\tilde{\Lambda}_{ p c}^{0}$ are real quantities.
Note that $q_f$ is independent of laser power since its numerator as
well as the denominator  is proportional to laser amplitude.
Following the Ref. \cite{verhaar}, we can express $\epsilon$ in terms of applied magnetic field
in the form
\bea
\epsilon = \frac{E - E_{th} - (E_{th} - \tilde{E}_f)}{\Gamma_f/2} =  \frac{E - E_{th}}{\Gamma_f/2} - \frac{B - B_0}
{\Delta (k a_{bg})}  \eea
where $E_{th}$ is the threshold of the open channel, $\Delta$ is the Feshbach resonance width, $B_0$ is the
resonance magnetic field and $a_{bg}$ is the background scattering length. Here $E - E_{th}$ is the asymptotic collision energy.  The energy $E_{th}$ depends on the applied magnetic field due to Zeeman shift of the atomic level. The resonance scattering length is given by $a_{res} = - a_{bg} \Delta/(B - B_0)$.
\begin{figure}
\includegraphics[width = 3.5 in]{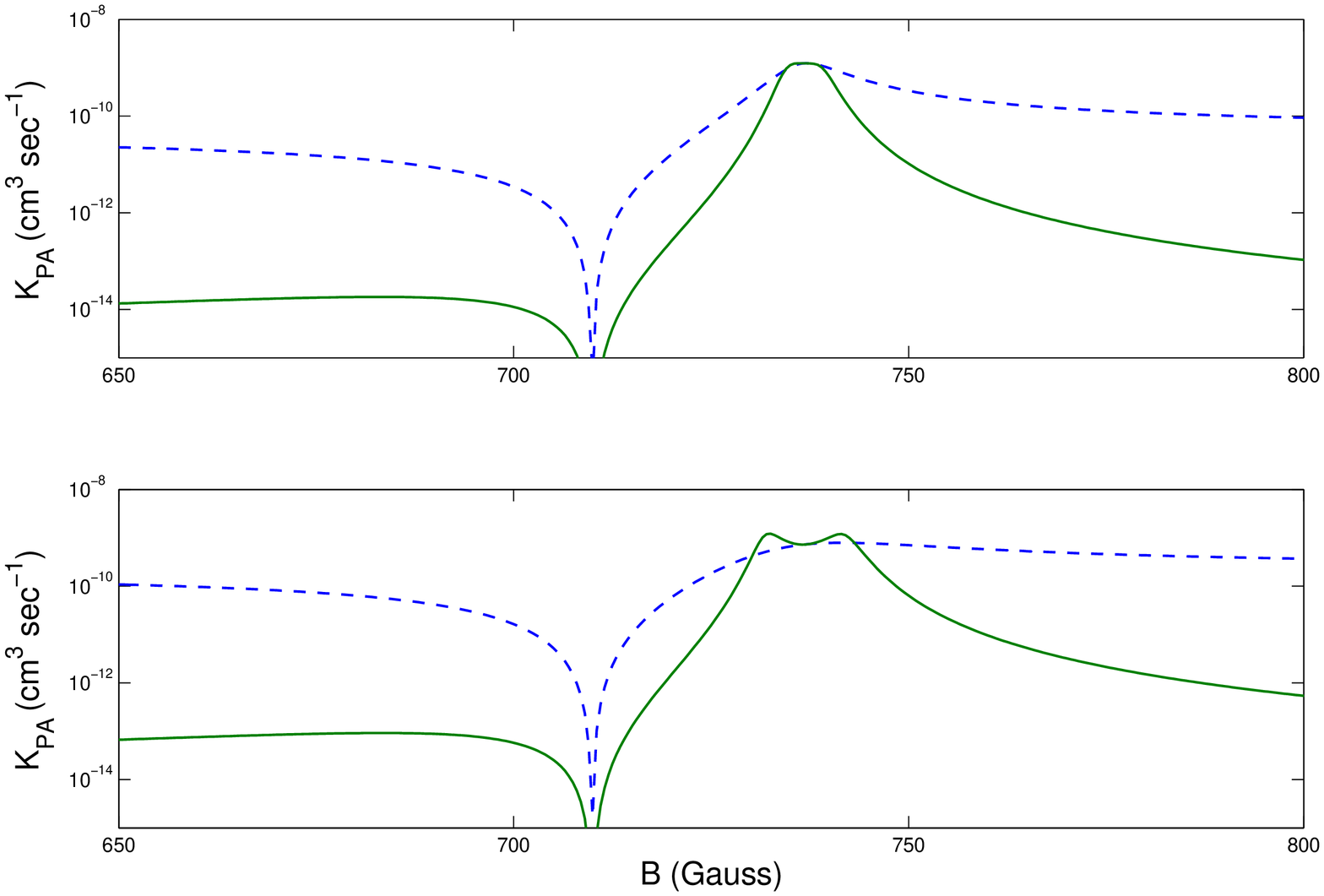}
\caption{The dashed lines are  $K_{PA}$ vs. $B$ plots  when PA laser is tuned on resonance with continuum-bound transition. The solid lines are corresponding plots when the laser is tuned on resonance with bound-bound transition. For upper panel
(a)  $\Gamma_p = 0.01 \Gamma_f$, and for lower panel (b) $\Gamma_p = 0.05 \Gamma_f$. For both the panels,  the other fixed  parameters are $q_{f} = -9.0$,  $\gamma = 0.7 \Gamma_f$,  $T = 10 \mu$K, $\Gamma_f = 16.67$ MHz and $B_0 = 736.72$ Gauss.
The
minimum and maximum of dashed curve in (a) appear at 710 Gauss and 736.8 Gauss, respectively.}
 \label{fig:2}
\end{figure}

Before we discuss our main results, we would like to point out how our mathematical treatment discussed above is related
to the recent work of Koznetsova {\it et al.}  \cite{Kuznetsova} who have studied a related model in a different context which is to  transfer of atoms into ground state molecules via two-photon process near a Feshbach resonance. Our approach is
to find out the real space dressed wave function  by solving time-independent scattering problem  by Green's function method  while they have adapted a quantum optics-based approach of finding time-dependent amplitudes of the dressed state by solving coupled differential equations numerically.

\section{The results and discussion}
We now discuss characteristic features of our main results.  PA  spectrum   is given by the PA rate coefficient $K_{PA} =  \langle  v_{rel} \sigma_{PA} \rangle$, where $\sigma_{PA} =  (\hbar \gamma |\tilde{A}_p|^2 )/(2 \pi k^2)$ is the cross section for the loss of atoms due to decay of the excited molecules. Here $\langle \cdots \rangle$ implies thermal averaging over
the relative velocity $v_{rel} = \hbar k /\mu$.  Note that, in
the limit $\Gamma_f \rightarrow 0$,  PA spectrum reduces to a
Lorentzian implying that coupling between closed channel bound state and the continuum is
essential for the occurrence of Fano interference. When both
$\Lambda_2$ and $\tilde{V}_{f c}^{0}$ go to zero, the spectrum
reduces to that of standard PA. For numerical illustrations,
we consider a model system of two ground-state (S$_{1/2}$) $^7$Li atoms undergoing PA from the ground molecular configuration $ {^3}\Sigma^+_u$ to the vibrational state $v = 83$ of the excited molecular configuration  ${^3}\Sigma_g$  which correlates asymptotically to 2S$_{1/2}$ + 2P$_{1/2}$ free atoms.  The spontaneous linewidth is taken to be $\gamma = 11.7$ MHz \cite{prodan}. The experimental value of shift $S_{pc}$ is reported be  $ -1.7 \pm 0.2$ MHz /W cm${^2}$ \cite{prodan}. The resonance width is $\Delta = - 192.3$ Gauss  and the background scattering length  $ a_{bg} =-24.5 a_0$ ($a_0$ is Bohr radius)  \cite{Pollack}. The Feshbach resonance linewidth $\Gamma_f$ at 10 $\mu$K temperature is calculated out to be 16.66 MHz using the parameters reported in Ref. \cite{chin}.

Depending on how PA laser is tuned, we have two cases. In the first case (case-I), laser is
on or near resonance with free-bound transition but off-resonant with bound-bound transition. In the second case (case-II), it is resonant with bound-bound transition.  PA rate will be maximized at the poles of Eq. (\ref{ap}). In case-I,  there is only one pole of Eq. (\ref{ap}) which depends on laser intensity. The minimum in the spectrum is solely determined by the asymmetry parameter $q_f$ and is  independent of laser intensity. We first consider case-I  and  plot  $(K_{PA})$ as a function of $B$ for three different values of  $\Gamma_p$ in Fig.2. For these three different $\Gamma_p$ values we choose three different $q_f$ parameters such that the maximum appears near $B = 736$ G \cite{Junker:prl:2008}.  Since the minimum position $B_{min}$ is independent of laser intensity, we also choose three different values of resonant magnetic field $B_0$ such that $B_{min}$ remains fixed at 710 G for theses three $\Gamma_p$ values. In this case there arises asymmetric Fano profile with one minimum and one maximum. This results from quantum interference between  continuum-bound and bound-bound Raman-type transition pathways. This interpretation of Raman Fano profile is in accordance with the recent experimental observation of two-photon PA by Moal {\it et al.} \cite{moal}.

Next we consider case-II in which PA laser is tuned in resonance with bound-bound rather than continuum-bound transition. In this case we have
$\hbar \delta - E_p + \tilde{E}_f = 0$ and so $\Delta_p = \epsilon$. Then $\tilde{A}_p$ will have two maxima given by
\bea
\epsilon (\epsilon + i) + i \tilde{\Gamma}_{t} ( \epsilon + i ) - \tilde{\Gamma}_{p} (q_f- i)^2 = 0.
\eea
where $\tilde{\Gamma}_t = \Gamma_t/\Gamma_f$ with $\Gamma_t = \Gamma_p + \gamma$ being the total line width, and $\tilde{\Gamma}_p = \Gamma_p/\Gamma_f$. For the sake of comparison, we plot spectra in Fig.3 for both the cases. Figure 3(a) shows that a single  maximum  appears in both the cases when laser intensities are low. As laser intensity increases, the maximum in case-I disappears while a two peak structure emerges in case-II as displayed in Fig.3(b). We notice that  PA rate is lower in case-II in comparison to case-I for the same magnetic field and other parameters except near the two maxima.
To further investigate into the double-peak structure, we demonstrate spectra for case-II at higher intensities in Fig.4 which clearly indicates the nonlinear features of Fano interference. The origin of the two peaks lies in Autler-Townes splitting \cite{eberly} due to bound-bound resonant coupling when continuum-bound dipole coupling is scanned into two-photon resonance.
At lower intensities, the two peaks can appear on the same side of Fano minimum. As a result, there can appear another smaller minimum (we call it Autler-Townes (AU) minimum in order to
distinguish it from Fano minimum) between the two peaks. By comparing Fig.3(a) with Fig.3(b), we note that the AU minimum at a higher intensity appears near the position where maximum would have appeared at a lower intensity.
The separation between the two maxima increases with increasing laser intensities as shown in Fig.4. As one of the peaks crosses the Fano minimum at an increased laser intensity, the AU minimum disappears due to its interference with the much stronger Fano minimum resulting in two-maximum structure only. The
double-maximum structure is particularly prominent in the
strong-coupling regime where  $\Gamma_p$  exceeds
the spontaneous line width $\gamma$ of PA
molecule. Recently, Pellegrini and Cote \cite{pellegrini:njp:2009} have theoretically obtained double-minimum spectra using the formalism of Ref. \cite{cote2} which is to first diagonalize the part of the Hamiltonian pertaining to the ground state scattering (continuum interacting with the bound state in closed-channel), and then to calculate the optical transition matrix element between this diagonalized state and the excited molecular state by Fermi-Golden rule. This is similar to linear Fano theory \cite{Fano61} and hence can not be applied for strong-coupling that can further modify the continuum state significantly. In our formalism, we have diagonalized the full Hamiltonian nonperturbatively.

\begin{figure}
\includegraphics[width = 3.5 in]{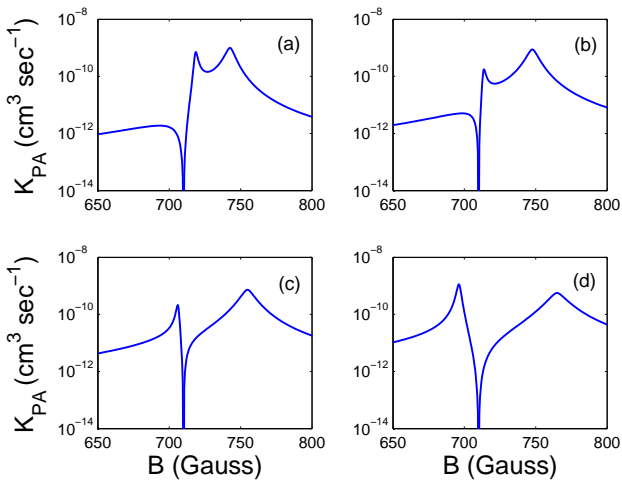}
\caption{Shown are $K_{PA}$ vs. $B$ plots  when PA laser is tuned on resonance with bound-bound transition
for $\Gamma_p = 0.5 \gamma$ (a),  $\Gamma_p = \gamma$ (b),  $\Gamma_p = 2 \gamma$ (c) and  $\Gamma_p = 4 \gamma$ (d).
The parameters chosen are $B_0 = 730.51$Gauss, $q_f = -6.89$ with all others parameters remaining same as in Fig.2}
 \label{fig:3}
\end{figure}

\section{conclusion}
 The results discussed above
clearly demonstrate linear and nonlinear aspects of Fano
interference in weak- and strong-coupling regimes, respectively.  Observation of two-minimum and
two-maximum structures crucially depends on precise tuning of PA laser on or near resonance with bound-bound transition.  If the laser field is tuned to get the maximum amount of loss of atoms for a fixed magnetic field, the resulting PA spectrum will mostly correspond to the case-I with a single maximum. To explore the nonlinear Fano effect, it is important to know the binding energy of the closed-channel bound state  so that the laser can be accurately tuned near resonance with the bound-bound transition as the magnetic field is varied.  Recently, nonlinear Fano effect was observed in quantum dot \cite{Recentnaturepaper}.
Although Autler-Townes splitting has been recently demonstrated in two-photon PA  \cite{two-photon,moal}, it is yet to be observed in PA with a single laser beam in the presence of Feshbach resonance. Fano interference may further  be explored in photoassociation
between heteronuclear atoms such as Na  and Cs \cite{bigelow} or K
and Rb \cite{ni} which have broad magnetic Feshbach resonance and shorter ranged
excited potentials.  \\
\\
{\bf Acknowledgment}\\
We are thankful to N. Bigelow for
discussions. This work is supported by the NSF Grant No. PHYS 0653494.


\begin{thebibliography}{999}

\bibitem{Fano61} U. Fano, Phys. Rev. {\bf 124}, 1866 (1961).

\bibitem{eberly}
K. Rzazewski and J. H. Eberly, Phys. Rev. Lett.
{\bf 47}, 408 (1981).

\bibitem{gsafano}  P. Lambropoulos and P. Zoller,
Phys. Rev. A {\bf 24}, 379 (1981); G. S. Agarwal {\it et al.}, Phys. Rev. Lett. {\bf 48}, 1164 (1982); G. S.
Agarwal, S. L. Haan,  and J. Cooper, Phys. Rev. A {\bf 29}, 2552
(1984).

\bibitem{EITHarrisPRL89} S. E. Harris , Phys. Rev. Lett. {\bf 62}, 1033 (1989); S. Harris, Physics Today {\bf 50}, 36
(1997); M. D. Lukin and A. Imamoglu, Nature
{\bf 413}, 273 (2001).

\bibitem{viiGSAbook} G. S. Agarwal, \textit{ Quantum Statistical Theories of Spontaneous
Emission and Their Relation to Other Approaches},
 Springer Tracts in Modern Physics: Quantum Optics (Springer-Verlag, Berlin, 1974);
 A. A. Svidzinsky, J. Chang, and M. O. Scully,
Phys. Rev. Lett. {\bf 100}, 160504 (2008).

\bibitem{scullyzhuzubairyScience} M. O. Scully and M. S. Zubairy,
Science {\bf 301}, 181 (2003).

\bibitem{HauHarris} L. V. Hau {\it et al.},  Nature {\bf 397}, 594 (1999).


\bibitem{recentpaperBoyd} Z. Shi {\it et al.}, Phys. Rev. Lett. {\bf 99}, 240801
(2007).

\bibitem{atom-molecule} R. Wynar {\it et al.}, Science {\bf 287}, 1016 (2000); K. Winkler {\it et al.}, Phys. Rev. Lett. {\bf 95}, 063202 (2005); C. Ryu {\it et al.}, cond-mat/0508201.

\bibitem{two-photon} R. Dumke {\it et al.}, Phys. Rev. A {\bf 72}, 041801(R) (2005).

\bibitem{moal} S. Moal {\it et al.} Phys. Rev.  Lett. {\bf 96}, 023203 (2006).

\bibitem{semian} J. L. Bohn and P. S. Julienne, Phys. Rev. A {\bf 60}, 414 (1999).

\bibitem{semian2} J. L. Bohn and P. S. Julienne, Phys. Rev. A {\bf 54}, R4637 (1996).

\bibitem{Junker:prl:2008} M. Junker { \it et al.}, Phys. Rev. Lett. {\bf 101}, 060406 (2008).

\bibitem{Winkler} K. Winkler { \it et al.}, Phys. Rev. Lett. {\bf 98},043201 (2007).

\bibitem{ni} K. K. Ni { \it et al.}, Science {\bf 322}, 231 (2008).

\bibitem{Mackie:prl:2008} M. Mackie { \it et al.}, Phys. Rev. Lett. {\bf 101}, 040401 (2008).

\bibitem{pellegrini:njp:2009} P. Pellegrini and R. Cote,  New J. Phys.{\bf 11}, 055047 (2009).

\bibitem{cote2} P. Pellegrini, M. Gacesa and R. Cote,  Phys. Rev. Lett. {\bf 101}, 053201 (2008).

\bibitem{Kuznetsova} E. Kuznetsova {\it et al.}, New J. Phys. {\bf 11}, 055028 (2009).

\bibitem{mfr} E. Tiesinga, B. J. Verhaar and H. T. C. Stoof, Phys. Rev. A {\bf 47},4114 (1993);
S. Inouye {\it et al.},
Nature {\bf 392}, 151 (1998);
 Ph. Courteille {\it et al.}, Phys. Rev. Lett. {\bf 81}, 69
(1998); J. L. Roberts {\it et al.}, Phys. Rev. Lett. {\bf 81},
5109 (1998).

\bibitem{pa} H. R. Thorsheim, J. Weiner, and P.S. Julienne,
Phys. Rev. Lett. {\bf 58}, 2420 (1987); for  reviews on PA, see  J. Weiner {\it et al.}, Rev. Mod. Phys. {\bf 71}, 1 (1999);
F. Masnou-Seeuws and P. Pillet, Adv. At. Mol.
Phys. {\bf 47}, 53(2001); K. M. Jones {\it et
al.}, Rev. Mod. Phys. {\bf 78}, 483 (2006).

\bibitem{verhaar} A. J. Moerdijk, B. J. Verhaar and A. Axelsson, Phys. Rev. A {\bf 51}, 4852 (1995).

\bibitem{prodan} I. D. Prodan { \it et al.}, Phys. Rev. Lett.{\bf 91} 080402 (2003).

\bibitem{Pollack} S. E. Pollack { \it et al.}, Phys.Rev. Lett.{\bf 102}, 090402 (2009).

\bibitem{chin} C. Chin {\it et al.} LANL arXiv 0812.1496 (2008).

\bibitem{Recentnaturepaper} M. Kroner {\it et al.}, Nature {\bf
451}, 311 (2008).

\bibitem{bigelow} J. P. Shaffer, W. Chalupczak, and N. P. Bigelow,
Phys. Rev. Lett. {\bf 82}, 1124 (1999); C. Haimberger {\it et al.}, Phys. Rev. A {\bf
70}, 021402(R) (2004).


\end{thebibliography}
\end{document}